      [1999/12/01 v1.4c Il Nuovo Cimento]
\documentclass{cimento}

\usepackage{graphicx}  
\title{QCD resummation in hadron production}
\author{D.~Anderle\from{ins:Tue}, F.~Ringer\from{ins:Tue}, W.~Vogelsang\from{ins:Tue}}

\instlist{ \inst{ins:Tue} Institute for Theoretical Physics, 
T\"ubingen University, Auf der Morgenstelle 14, 72076 T\"ubingen, Germany}

\PACSes{\PACSit{13.85.Ni, 12.38.Bx}.}
\begin{document}

\maketitle

\begin{abstract}
We present calculations of next-to-leading order and resummed QCD corrections for semi-inclusive deep-inelastic scattering and single-inclusive $e^+e^-$ annihilation. The resummation is performed to next-to-leading logarithmic accuracy. Knowing these QCD corrections is important in order to extract parton distribution functions and fragmentation functions from present and future data for these
reactions. We present comparisons of our results to recent data by
the COMPASS, HERMES, Belle, and BaBar experiments.
\end{abstract}

\section{Introduction \label{intro}}

Processes with identified final-state hadrons play important roles in QCD.  In the present work, we address higher-order perturbative corrections to two of the key hadron production processes, single-inclusive annihilation (SIA) $e^+e^-\to hX$ and semi-inclusive deep-inelastic scattering (SIDIS),  $\ell p\to \ell hX$, where  $h$ denotes a final-state hadron. Modern analyses~\cite{defloriandss,akk,hkns} of fragmentation functions variously use data for these two processes. Our study is very much motivated by the recent advent of data for these reactions with unprecedented high precision. The Belle collaboration at 
KEK~\cite{ref:belleresults}, and BaBar at SLAC~\cite{Lees:2013rqd}
have presented data for pion and kaon multiplicities in SIA with a very fine binning and extremely high precision at the sub-1\% level. New preliminary
high-statistics SIDIS data have been shown by the HERMES~\cite{hermes} and COMPASS~\cite{compass} lepton scattering experiments over the past year or so. 

In the kinematic regimes accessed by these experiments, perturbative-QCD corrections are expected to be fairly significant. The infrared cancellations between virtual and real-emission diagrams in higher orders leave behind logarithmic contributions. These ``threshold logarithms'' become large when the phase space for real-gluon radiation is shrinking and therefore have to be taken into account to all orders in perturbation theory. In this work, we will examine 
this resummation of the large corrections. We will
restrict ourselves to resummation at next-to-leading logarithmic (NLL) accuracy, which should capture the main effects.
A more detailed description of the analysis presented in this paper was published in ref.~\cite{ref:anderle:2012rq}.

\section{Resummation for SIDIS multiplicities \label{sidis}}

We consider semi-inclusive deep-inelastic scattering $\ell p\to \ell hX$. Using the usual kinematic variables, the SIDIS cross section may be written as~\cite{altarelli}
\begin{equation}\label{eq:sidis}
\frac{d^3\sigma^h}{dx dydz} = \frac{4\, \pi\alpha^2}{Q^2} \left[ \frac{1+(1-y)^2}{2y} {\cal F}_T^h(x,z,Q^2)+ \frac{1-y}{y} {\cal F}_L^h(x,z,Q^2) \right],
\end{equation}
where $\alpha$ is the fine structure constant and ${\cal F}_{T,L}^h$ are the transverse and the longitudinal structure functions. SIDIS hadron multiplicities are defined by 
\begin{equation}\label{Rdef}
R^h_{\mathrm{SIDIS}}\equiv\frac{d^3\sigma^h/dx dydz}{d^2\sigma/dxdy},
\end{equation}
where $d^2\sigma/dxdy$ is the cross section for inclusive DIS, $\ell p\to \ell X$. In order to investigate higher-order effects on SIDIS multiplicities, we have to consider QCD corrections to both the SIDIS and the inclusive DIS cross section. 

Using factorization, the transverse and longitudinal structure functions in eq.~(\ref{eq:sidis}) are 
given by ($i=T,L$)
\begin{equation}\label{eq:f1sidis}
\hspace*{4mm}{\cal F}_i^h(x,z,Q^2) =\sum_{f,f'} \int_x^1 \frac{d\hat{x}}{\hat{x}}\int_z^1 \frac{d\hat{z}}{\hat{z}}  f \left(\frac{x}{\hat{x}},
\mu^2\right) D^h_{f'} \left(\frac{z}{\hat{z}},\mu^2\right) \,{\cal{C}}^i_{f'f}
\left(\hat{x},\hat{z},\frac{Q^2}{\mu^2},\alpha_s(\mu^2)\right) , 
\end{equation}
where $f(\xi,\mu^2)$ denotes the distribution of parton $f=q,\bar{q},g$ in the nucleon at momentum fraction $\xi$ and scale $\mu$, while $D^h_{f'} \left(\zeta,\mu^2\right)$ is the corresponding fragmentation function for parton $f'$ going to the observed hadron $h$. The hard-scattering coefficient functions ${\cal{C}}^i_{f'f}$ can be computed in perturbation theory. 

Since threshold resummation can be derived in Mellin-moment space~\cite{dyresum1,dyresum2}, it is useful to take Mellin moments of the structure functions ${\cal F}_T^h$ and ${\cal F}_L^h$. Since $x$ and $z$ are independent variables, we take moments separately in both. We find from eq.~(\ref{eq:f1sidis})
\begin{equation}\label{Melmo}
\tilde{{\cal F}}^h_i(N,M,Q^2)=\sum_{f,f'}\tilde{f}^N(\mu^2)\tilde{D}_{f'}^{h,M}(\mu^2)\,\tilde{{\cal C}}^i_{f'f}\left(N,M,\frac{Q^2}{\mu^2},\alpha_s(\mu^2)\right).
\end{equation}
Thus, the Mellin moments of the structure functions are obtained from ordinary products of the moments of the parton distribution functions and fragmentation functions, and double-Mellin moments of  the partonic hard-scattering functions. 
To NLO, the latter may be found in~\cite{sv}.
Using the techniques developed in ref.~\cite{ref:Sterman:2006hu}, we find that the final NLL resummed coefficient function becomes in the $\overline{\mathrm{MS}}$ scheme (see also~\cite{cacc})
\begin{eqnarray}\label{resummed3}
\hspace*{6mm}{\cal{C}}^{T,{\mathrm{res}}}_{qq}(N,M,\alpha_s(Q^2))&=&e_q^2 H_{qq}\left(\alpha_s(Q^2),\frac{Q^2}{\mu^2}\right)
\exp\left[2 \int_0^{Q^2}  \frac{dk_\perp^2}{k_\perp^2}A_q\left(\alpha_s(k_\perp^2)\right) \right. \nonumber \\
&& \left. \times \left\{ K_0 \left(\sqrt{NM}\,\frac{2k_{\perp}}{Q} \right)+  \ln \left(\frac{k_{\perp}}{Q}\sqrt{\bar{N} \bar{M}}\right) \right\}\right].
\end{eqnarray}
Here $A_q$ and $H_{qq}$ are perturbative functions. For resummation at NLL, one needs to know $H_{qq}$ to first order in the strong coupling collecting the hard virtual corrections. 

As the exponentiation of soft-gluon corrections is achieved in Mellin moment space, the hadronic structure function is obtained by taking the inverse Mellin transforms of eq.~(\ref{Melmo}), which are given by contour integrals in the complex plane.

The NLL resummed results for DIS and SIA can be found in ref.~\cite{ref:anderle:2012rq} and refs. therein.

\section{Phenomenological Results \label{Pheno}}

We now investigate the numerical size of the threshold resummation effects for SIDIS and SIA multiplicities. For the parton distribution functions we use the NLO MSTW 2008 set of~\cite{Martin:2009iq}, whereas we choose the NLO DSS~\cite{defloriandss} pion fragmentation functions.

Starting with SIDIS, fig.~\ref{ref:fig1} (\ref{ref:fig2}) shows our results for the NLO and the resummed multiplicities with COMPASS (HERMES) kinematics. The center-of-mass energy of COMPASS is $\sqrt{s}\approx$17.4 GeV. The kinematic cuts employed are $0.041<x<0.7$, $0.1<y<0.9$, $Q^2>1$ GeV${}^2$ and $W^2=Q^2(1-x)/x+m_p^2>49$ GeV${}^2$, where $m_p$ is the proton mass.  For HERMES the center-of-mass energy is $\sqrt{s}\approx$7.26 GeV. The kinematic cuts are $0.023<x<0.6$, $0.1<y<0.85$, $Q^2>1$ GeV${}^2$ and $W^2>10$ GeV${}^2$. We choose the renormalization and factorization scales as $Q$ and consider the SIDIS multiplicity for $\pi^-$ as a function of $z$, but integrating numerator and denominator of eq.~(\ref{Rdef}) over $x$ and $y$. It turns out that resummation leads to a moderate, but significant, enhancement of the multiplicities.

In fig.~\ref{ref:fig3}, we present our results for the SIA $\pi^-$ multiplicity at $\sqrt{s}=10.52$~GeV and for $-1<\cos\theta<1$. The results can also be studied as ratios $({\mathrm{Th' - NLO}})/{\mathrm{NLO}}$, where ${\mathrm{Th'}}$ denotes any of the higher-order SIDIS multiplicities generated by resummation. These ratios are shown in fig.~\ref{ref:fig4}, making the large resummation effects even more apparent at high $x_E$. One observes that the new high-precision
data from Belle~\cite{ref:belleresults} and BaBar~\cite{Lees:2013rqd} are
well consistent with enhancements predicted by resummation.


\begin{figure}[h!]
 Ê\hfill
 Ê\begin{minipage}[t]{.45\textwidth}
 Ê Ê\begin{center} Ê
 Ê Ê \includegraphics[width=.97\textwidth,angle=90]{Felix_Ringer_QCDN12_figure1.epsi}
    \vspace*{-0.5cm}
 Ê Ê Ê\caption{\label{ref:fig1} NLO and NLL resummed SIDIS multiplicities for $\pi^-$ compared to the preliminary COMPASS data~\cite{compass}.}
 Ê Ê\end{center}
 Ê\end{minipage}
 Ê\hfill
 Ê\begin{minipage}[t]{.45\textwidth}
 Ê Ê\begin{center} Ê
 Ê Ê\includegraphics[width=.97\textwidth,angle=90]{Felix_Ringer_QCDN12_figure2.epsi}
        \vspace*{-0.5cm}
 Ê Ê Ê\caption{\label{ref:fig2} Same as Fig.~\ref{ref:fig1}, but for HERMES kinematics. The preliminary
                      data are from~\cite{hermes}. }
 Ê Ê\end{center}
 Ê\end{minipage}
 Ê\hfill
\end{figure}



\begin{figure}[h!]
 Ê\hfill

 Ê\begin{minipage}[t]{.45\textwidth}
 Ê Ê\begin{center} Ê
    \vspace{-0.2cm}
 Ê Ê\includegraphics[width=0.96\textwidth,angle=90]{Felix_Ringer_QCDN12_figure31.epsi}
        \vspace*{-0.38cm}
 Ê Ê Ê\caption{$(\pi^++\pi^-)/2$ multiplicity in $e^+e^-$ annihilation computed 
at $\sqrt{s}=10.52$~GeV. 
The data are from Belle~\cite{ref:belleresults} and BaBar~\cite{Lees:2013rqd} (``conventional'' data set, $z\geq 0.2$, $\sqrt{s}=10.54$~GeV).}
 Ê Ê Ê\label{ref:fig3}
 Ê Ê\end{center}
 Ê\end{minipage}
 Ê\hfill
Ê\begin{minipage}[t]{.45\textwidth}
 Ê Ê\begin{center} Ê
 Ê Ê \includegraphics[width=0.9\textwidth,angle=90]{Felix_Ringer_QCDN12_figure4.epsi}
    \vspace*{-0.5cm}
 Ê Ê Ê\caption{Ratios $({\mathrm{Th' - NLO}})/{\mathrm{NLO}}$, where ${\mathrm{Th'}}$ corresponds to the 			   multiplicity at higher orders as generated by resummation.}
\label{ref:fig4} 
 Ê Ê\end{center}
 Ê\end{minipage}
 Ê\hfill
\end{figure}


\vspace*{-18mm}
\section{Conclusions \label{sum}}

We have derived threshold-resummed expressions for the coefficient functions for the SIDIS and SIA processes. We have presented comparisons to pion 
multiplicities measured in these processes 
at COMPASS, HERMES, Belle, and BaBar. We found that resummation leads to modest but significant enhancements of the multiplicities. Our main point is that, given the good accuracy of the new preliminary data sets,  it will be crucial to include resummation effects for both processes in the next generation of analyses of fragmentation functions.


\end{document}